\magnification 1000
\def\mincir{\raise -2.truept\hbox{\rlap{\hbox{$\sim$}}\raise5.truept
\hbox{$<$}\ }}
\def  \magcir{\raise -2.truept\hbox{\rlap{\hbox{$\sim$}}\raise5.truept
\hbox{$>$}\ }}
\def\rn{$ ~{0.06 \, h_{50}^{-2} \over \Omega_{BBN} }~$}
\def\ref{\par\noindent\hangindent 20pt}

\centerline{\bf DARK BARYONS IN THE UNIVERSE:} 
\centerline{\bf THE QUEST GOES ON}
\bigskip
\centerline{\bf Paolo Salucci$^{1}$ and Massimo Persic$^{1,2}$}
\medskip
\centerline{$^1$ SISSA -- International School for Advanced Studies,
via Beirut 2-4, I--34013 Trieste, Italy}
\centerline{$^2$ Osservatorio Astronomico, via G.B. Tiepolo 11, I-34131
Trieste, Italy}
\centerline{salucci@galileo.sissa.it, persic@sissa.it}
\bigskip
\bigskip

\noindent
{\bf Abstract.} We show that the high local baryonic fraction, $M_{bar} \sim
1/3\, M_{tot}$, 
found in groups and clusters of galaxies does not reconcile the observed
cosmological baryon density with the standard Big-Bang prediction. Taking into 
account recent measurements on the hot-gas content and temperature functions 
of clusters and groups, we get $\Omega_{gas}^{cg} ~= ~2.3 \times 10^{-3}h_{50}^
{-1.5} \simeq 4 \% \,\Omega_{BBN}$. Including the contributions of galaxies 
and of (local) Ly$\alpha$ clouds we estimate $\Omega_{bar} \sim (4-6)\times 
10^{-3} \mincir 10\% \, \Omega_{BBN}$ as the amount of detected baryons. The 
great majority of the synthesised atoms are still to be discovered. We propose 
to relate the impressive presence of the hot gas component in clusters with 
the very low, mass-dependent efficiency of the process of galaxy formation in 
making stars from the primordial gas. 
\bigskip

\leftline{\bf Introduction: The Baryon Overabundance in Clusters.}

In a previous paper (Persic \& Salucci 1992, hereafter PS92) we computed
the cosmological density of the baryonic matter, $\Omega_{bar}$, by means of an
inventory of the stellar and gaseous content of galaxies, groups, and clusters
(see also Bristow \& Phillips 1994). The value we found, $\Omega_{bar} \simeq
3^{+1}_ {-2} \times 10^{-3}$, implies that most of the cosmologically
synthesised baryons do not reside in luminous structures but are disguised
elsewhere in some invisible form. The above estimate relies on accurate 
dynamical methods able to derive the amount of luminous matter in the various 
cosmic structures. However, as far as groups and clusters of galaxies are 
concerned, their gas content 
$$
\Omega_{bar}^{cg}~ =~ {1\over \rho_c} \int^{M_{gas}^{max}}_{M_{gas}^{min}} 
n(M_{gas})\, M_{gas} \, dM_{gas}\,,
\eqno(1)
$$ 
which accounts for virtually the whole baryonic content of these systems, was
computed by: {\it (a)} a 
linear extrapolation of the inner IGM mass profiles out to the Abell radius, 
$R_A$; and {\it (b)} assuming that groups have the same gas-mass function of 
clusters extrapolated to lower masses/temperatures. The value obtained,
$\Omega_{bar}^{cg} = 1.5\times 10^{-3}h_{50}^{-1.5}$, was therefore somewhat 
uncertain, although definitely much smaller than the standard Big Bang 
Nucleosynthesis (hereafter BBN) prediction $\Omega_{BBN} \simeq 0.06 
h_{50}^{-2}$. In the past two years ROSAT and ASCA have mapped, in several 
clusters and groups, the density and temperature profiles of the intracluster 
(IC) gas out to $\sim R_A$ (e.g., Lubin et al. 1995), allowing 
the main structural properties of these structures to be established. A
roughly spherical 
gaseous halo, most likely in thermal equilibrium and showing small temperature 
gradients (i.e. with $T(r) \sim const$), surrounds virtually every cluster and 
group. The space density is obtained by deprojecting the surface brightness of 
the emitting gas $I(r)\propto [1+({r\over r_X})^2]^{-3\beta +1/2}$ (Cavaliere 
\& Fusco-Femiano 1976) to obtain: $\rho_X(r) ~\propto~ [1+ \bigl({ 
r\over r_X}\bigr)^2]^{-3 \beta/2}$;
the gravitating mass density is obtained by means of the 
hydrodynamic 
equilibrium equation (see Sarazin 1986): since 
$\beta \simeq {2/3}$ (e.g., Davis, Forman \& Jones 1995), it follows
$\rho(r)~\propto~[3-({r\over
r_X})^2] / [1+({r\over
r_X})^2]^2$. Therefore, the dark-to-baryonic density ratio decreases 
with radius so that in most clusters and groups the visible-to-dark mass ratio 
at $\sim R_A$ exceeds the cosmological limit $\Omega_{BBN}/ \Omega$ (unless
$\Omega$ is low) (e.g.: David et al. 1995; Dell'Antonio, Geller \& Fabricant 
1995; Pildis, Bregman \& Evrard 1995). In this paper we investigate whether 
such a high fraction is relevant for the quest of baryonic matter launched by 
PS92: can the IC gas, once properly taken into account, remove the 
need for baryonic dark matter (i.e., is $\Omega^{cg}_{bar} \simeq
\Omega_{BBN}$)? 

The main aim of this paper is to re-estimate $\Omega_{bar}^{cg}$ by means of
the actual cluster (and group) gas masses and temperature functions, and
to evaluate $\Omega_{bar}$ using also recent work on galaxy 
structure. A second aim is to explore a way out from the "baryon catastrophe"
in clusters, by showing how their high baryonic fraction may be a consequence 
of a tight link between the two baryonic components of clusters, i.e.
the gas and the galaxies.

A value of Hubble constant of 50 km s$^{-1}$ is assumed throughout the paper.
\bigskip

\leftline{\bf The Determination of $\Omega_{bar}^{cg}$}

In terms of the cluster temperature $T$ we have: 
$$
\Omega_{bar}^{cg}~=~ {1\over \rho_c} \int^{T_{max}}_{T_{min}} n(T)\,
M_{gas}(T) \,  dT \,.
\eqno(2)
$$
Let us now examine each of these functions in turn.
\bigskip

\noindent
$\bullet$ {\it Temperature Function.} Both clusters and groups are represented
by a single temperature function: 
$$
n(T)\, dT ~= ~2 \times 10^{-4} ~T^{-4.7}~ dT ~~~ {\rm Mpc}^{-3}\,.
\eqno(3)
$$
with $T$ in keV (see Henry \& Arnaud 1991, Henry et al. 1995). Notice that for
clusters ($T>3$ keV) eq.(3) is in substantial agreement with the Schechter-like 
function of Edge et al. (1990) used in PS92, while for groups ($T<3$ keV)
eq.(3) increases with decreasing temperature faster than the extrapolation of 
Edge et al. (1992) used in PS92.
\bigskip

\noindent
$\bullet$ {\it Gas Mass--Temperature Relation.} In Appendix A we show that
the gas content of clusters and groups relates with other structure properties; 
in particular: 
$$
~ ~ ~ ~ ~ ~ ~ ~ ~ ~ ~ M_{gas}~ =~ 8 \times 10^{12} T^{1.2} M_{\odot}\, ~ ~ ~
~ ~ ~ ~ ~ ~ ~ ~ ~ T \geq 3 keV 
\eqno(4a)
$$
$$
~ ~ ~ ~ ~ ~ ~ ~ ~ ~ ~ M_{gas}~ =~ 2 \times 10^{12} T^{2.5} M_{\odot}\, ~ ~ ~
~ ~ ~ ~ ~ ~ ~ ~ ~ T < 3 keV 
\eqno(4b)
$$
The temperatures and the gas masses range, respectively, between 1 keV $< kT
<$10 keV and $10^{12} M_\odot < M_{gas} < 5 \times 10^{13}M_\odot$.
Therefore, from  eqs.(2),(3),(4) we obtain:
$$
\Omega_{gas}^{cg} ~= ~2.3 \times 10^{-3}h_{50}^{-1.5} \simeq 3 \% \,
\Omega_{BBN} \,,
\eqno(5)
$$
i.e. $50\%$ bigger than the PS92 estimate. This increase reflects the larger 
number density of ROSAT groups with respect to that implied the temperature 
function adopted in PS92. 

Thus, clusters and groups, being rare events, contribute very little to the
present-day baryonic density, in spite of their high baryonic fraction.
Structures with $kT \mincir 1$ keV (such as galaxies, binaries, triplets, and
compact groups) have a negligible gas content.
\bigskip

\centerline{\bf The Estimate of $\Omega_{bar}$.}

\noindent
$\bullet$ {\it Galaxies.} We estimate the baryon content of E/S0 and
spiral galaxies as in PS92: 
$$
\Omega_{bar}^{E,S}~ =~ {1\over{\rho_c}} \int^{L_{min}}_{L_{max}}  
\biggl( {M\over 
L} \biggr)_*^{E,S} \Phi^{E,S}(L)~ L\, dL\,.
\eqno(6)
$$
We get (see Appendix B):
$$
\Omega_{bar}^{gal}~=~ \Omega_{bar}^E ~+~ \Omega_{bar}^S ~=~ 2 \times
10^{-3}\,, 
\eqno(7)
$$
in good agreement with PS92 and with Bristow \& Phillips (1994). Remarkably, 
spirals and ellipticals contribute the same cosmological stellar mass
density. 
\bigskip

\noindent
$\bullet$ {\it Ly$\alpha$ Clouds.} HST observations have revealed that the
most numerous structures in the universe belong to a population of absorbing 
clouds of $\sim 100$ kpc radius and $\sim 10^9 M_\odot$: although their actual 
number density and physical properties are uncertain, their contribution to 
$\Omega_{bar}$ may be comparable with $\Omega_{bar}^{gal}$ (Shull, Stocke \& 
Penton 1995). 
\medskip

We then estimate:
$$
 \Omega_{bar} ~=~ (2^{+0.3}_{-0.5} \ + 2^{+0.5}_{-1} \ h_{50}^{-1.5}
+2^{+1}_{-2}) \times 10^{-3} ~=~(4-6)\times 10^{-3} ~\mincir~ 10\% \, 
 \Omega_{BBN}\,. 
\eqno(8)
$$
\bigskip

\leftline{\bf Discussion}

We propose that the large amount of hot gas in clusters is the signature of the
inefficiency of the process of disk/spheroid formation in transforming the
primordial material into stars. The $N_{gal}$ member galaxies in a cluster are
distributed with luminosity according to a Schechter-type function:
$$ 
\phi(L) \, dL ~ \propto ~ L^{-\alpha } e^{-L/L_*} \,dL\,,
\eqno(9)
$$
with $\alpha \sim 2$ (e.g., De Propris et al. 1995). 
Given the definition of $n_{rc}$, the number of galaxies defining the Abell
richness class (Abell 1958), we find $N_{gal} \sim 100 ~n_{rc}$. Then, since 
$$
n_{rc} ~=~ 10 \times T^{1.2}
\eqno(10)
$$
(see Fig.1), we get:
\vglue 6.truecm

{\it Figure 1. The relation between Abell richness class and temperature (see 
Girardi et al. 1995)} 
\vglue 0.5truecm
$$
M_{gas} ~ \simeq ~ 10^{10} M_\odot \times  N_{gal}\,.
\eqno(11)
$$ 
Eq.(11) suggests that in clusters the total amount of gas per member galaxy
is 
independent of other cluster properties, and resembles a universal constant. 
In Persic, Salucci \& Stel (1995; hereafter PSS95) we have shown that spiral
proto-halos with pre-virialization gas-to-dark mass ratio $\Omega_{BBN}/
{\Omega}$ turn only a fraction 
$$
f~ =~ 0.65 ~(L/L_*)^{-0.6}
\eqno(12)
$$
of their original baryon content into stars. Ellipticals show a similar
behaviour (Bertola et al. 1993). Therefore, the amount of baryonic material 
left unused by the process of forming the disks and spheroids of the $N_{gal}$ 
cluster galaxies is:
$$
0.06 \int M_*(L) ~ [1-f(L)] ~ \phi(L)\, dL~ = ~10^{10}\, M_\odot \times 
N_{gal}\,,
\eqno(13)
$$
where $M_*(L)$ is the stellar mass of a galaxy of luminosity $L$ (see PSS95 
and Bertola et al. 1993). This amount is comparable with the  
mass of the IC gas given by eq.(11): most of the 
IC medium has never been processed by stars, in agreement with cluster chemical
abundances that independently imply that a large fraction of the ICM is
primordial (Gibson \& Matteucci 1995). 

We suggest the following scenario: clusters switch on the ionised gas, unused 
in the disk/spheroid formation in cluster environment, by providing the 
(gravitational) potential needed to heat this medium up to the ambient virial 
temperature ($\sim 5$ keV). The gas left over in the formation
of field galaxies, on the other hand, wanders invisible in the cosmic void.

\leftline{\bf Conclusions}

Of all the atoms synthesised in the Big Bang:
\medskip

\noindent
$\bullet ~~~$  3\% \rn are locked up in stars;

\noindent 
$\bullet ~~~$  4\% \rn form the ICM;

\noindent 
$\bullet ~~~$  3\% \rn form the local population of Ly$\alpha$ clouds,
\medskip

\noindent 
so that, we still have to discover about (100 $-$ 10 $\times $ \rn)\% of the 
primordially synthesised baryons. Let us notice that also in the light of a 
low baryon density, $\Omega_{BBN} \simeq 2.5 \times 10^{-2}h_{50}^{-2}$ (see 
Hogan 1995), the fraction of missing baryons is still $\magcir 50\%$. 

Where are they? We suggest one ore more of the following: 
 
\medskip

\noindent
$\bullet$ {\it Diffuse Ionised IGM.} It naturally fits into the proposed 
scenario of primordial gas leftover from galaxy formation and remaining unseen
if spread out in the field.
\medskip

\noindent
$\bullet$ {\it Halo Jupiters.} The detection of MACHOs in the Galactic halo
(Alcock et al. 1995) may have revealed some baryonic dark matter, in the form 
of substellar objects as allowed by the {\it local} dynamical $M/L$ ratios in 
spiral galaxies (PSS95).
\medskip

\noindent
$\bullet$ {\it Faint Galaxies.} A large population of low-surface-brightness 
dwarfs, detected at low flux limits, $B \sim 24.0$ (Driver et al. 1994), are 
missing from the much brighter samples used to derive galaxy LFs (see McGaugh 
1995), and may contain a cosmologically relevant amount of baryons. 

\bigskip
\bigskip
\centerline {\bf References}
\vglue 0.2truecm

\ref{Abell, G.O. 1958, ApJS, 3, 211}
\ref{Alcock, C., et al. 1995, Phys. Rev. Lett., 74, 2867}
\ref{Bertola, F., Pizzella, A., Persic, M., \& Salucci, P. 1993, ApJ, 416, L45}
\ref{Bristow, P.D., \& Phillips, S. 1994, MNRAS, 267, 13}
\ref{Carr, B.J. 1994, ARA\&A, 32, 531}
\ref{Cavaliere, A., \& Fusco-Femiano, R. 1976, A\&A, 49, 137}
\ref{David, L.P., Jones, C., \& Forman, W. 1995, ApJ, 445, 578}
\ref{De Propris, R., Pritchet, C.J., Harris, W.E., McClure, R.D. 1995, ApJ, 
     450, 534}
\ref{Dell'Antonio, I.P., Geller, M.J., \& Fabricant, D.G. 1995, AJ, 110, 502}
\ref{Driver, S., Phillips, S., Davies, J., Morgan, I., \& Disney, M. 1994,
     MNRAS, 268, 393}
\ref{Edge, A.C., Stewart, G.C., Fabian, A.C., \& Arnaud, K.A. 1990, 
     MNRAS, 245, 559}
\ref{Franceschini, A., Danese, L., Granato, G.L., Mazzei, P., 1995:
    {\it Infrared galaxy evolution and contribution to the background flux},
    in {\it Unveiling the Cosmic Infrared Backgound}, ed. E.Dwek et al.
    (College Park, MD), in press.}
\ref{Gibson, B.K., \& Matteucci, F. 1995, MNRAS, submitted}
\ref{Girardi, M., Fadda, D., Giuricin, G., Mardirossian, F., Mezzetti, M.,
    \& Biviano, A. 1995, ApJ, in press}
\ref{Henry, J.P. \& Arnaud, K.A. 1991, ApJ, 372, 410}
\ref{Henry, J.P., Gioia, I.M., Huchra, J.P., Burg, R., McLean, B., 
    B\"ohringer, H., Bower, R.G., Briel, U.G., Voges, W., MacGillivray, H., 
    \& Cruddace, R.G. 1995, ApJ, 449, 422}
\ref{Hogan, C.J. 1995, astro-ph/9512003}
\ref{Lubin, L.M., Cen, R., Bahcall, N.A., \& Ostriker, J.P. 1995, ApJ, in
     press}
\ref{McGaugh, S.S. 1995, MNRAS, in press (astro-ph/9511010)}
\ref{Persic, M., \& Salucci, P. 1992, MNRAS, 258, 14P (PS92)}
\ref{Persic, M., Salucci, P., \& Stel, F. 1995, MNRAS, in press (PSS95)}
\ref{Pildis, R.A., Bregman, J.N., \& Evrard, A.E. 1995, ApJ, 443, 514}
\ref{Sarazin, C.S. 1986, Rev. Mod. Phys., 58, 1}
\ref{Shull, J.M., Stocke, J.T., \& Penton, S. 1995, AJ, in press
     (astro-ph/9510054)}
\ref{van der Marel, R.P. 1991, MNRAS, 253, 710}

\bigskip
\bigskip
\bigskip

\centerline{\bf APPENDIX A}
\bigskip

\noindent
In rich clusters, the dynamical mass and the gas mass are related by:
$$
M_{gas} ~=~ 0.13~ G^{-1} \sigma^2 R_A
\eqno(A1)
$$
(see Lubin et al. 1995), with $\sigma$ the line-of-sight velocity dispersion
of the galaxies. By means of the well-known $\sigma$--$T$ relationship: 
$$
\sigma~ =~340 ~  T^{0.6}~  {\rm km ~s}^{-1}
\eqno(A2)
$$
(see Girardi et al. 1995), we get eq.(4a). Using recent data, we plot the 
$M_{gas}$--$T$ data also for a number of poor clusters and groups (see Fig.2), 
down to 1 keV, whose fit is shown in eq.(4b).

\vglue 6.truecm

{\it Figure 2. The gas mass versus temperature relation. The flatter
portion of the 
curve is the relationship for clusters given by Lubin et al. (1995). The 
steepest part is our fit for the poor clusters and loose groups (with data 
 from Dell'Antonio et al. 1995). (Compact groups have a significantly lower
gas content, see Pildis et al. 1995.)}  
\vglue 0.5truecm

\bigskip
\centerline{\bf APPENDIX B} 
\bigskip

\noindent
The luminosity functions of galaxies take the usual Schechter form $h=1/2$):
$$
\phi(L) \, dL~=~ \phi_* ~ \biggl({L \over L_*}\biggr)^{-\alpha} ~ e^{-L/ L_*}
 ~{dL \over L_*} \,.
\eqno(B1)
$$
The $B$-band LF parameters are shown in Table 1. The $B$-band stellar
$(M/L)_*$ ratios scale with luminosity as 
$$
\biggl( {M_* \over {L}}\biggr)_{E/S0}~=~4~ \biggl({L \over L_*}\biggr)^{0.35}
\eqno(B2)
$$
$$
\biggl( {M_* \over L}\biggr)_S ~=~
2 \times {\rm dexp} \biggl[0.35~ {\rm log} \biggl({L \over
L_*}\biggr) - 0.75 ~ {\rm log}^2 \biggl({L \over L_*}\biggr) \biggr]\,, 
\eqno(B3)
$$
for elliptical and spiral galaxies, respectively (see van der Marel 1991
and PSS95, respectively). We follow the PS92 procedure. Notice that here we 
adopt: {\it (1)} the luminosity function for ellipticals as derived from a 
recent analysis of several cumulative LFs (Franceschini et al. 1995); and {\it 
(2)} the spiral disks' $M/L$ ratios as derived from mass modelling of about 
1100 rotation curves (PSS95). Notice that in both cases there is no substantial 
difference from the functions adopted in PS92.

\vglue 0.5truecm
\centerline{\it ... Table 1. ...}

\bye